# Microscopic imaging of non-repetitive dynamic scenes at 5 THz frame rates by time and spatial frequency multiplexing

Jungho Moon[1,2,+], Seok-Chan Yoon[1,2,+], Yong-Sik Lim[3], and Wonshik Choi[1,2,*]

[1]*Center for Molecular Spectroscopy and Dynamics, Institute for Basic Science, Seoul 02841, Korea*
[2]*Department of Physics, Korea University, Seoul 02841, Korea*
[3]*Department of Nano Science and Mechanical Engineering and Nanotechnology Research Center, Konkuk University, Chungbuk 380-701, Korea*
*mooon936@gmail.com*, *sdlkfwpeorup@gmail.com*, *yslim@kku.ac.kr*, [*]*wonshik@korea.ac.kr*
[*]*+82-10-8758-4710*
+: These authors contributed equally to this work.

**Abstract**
Femtosecond-scale ultrafast imaging is an essential tool for visualizing ultrafast dynamics in molecular biology, physical chemistry, atomic physics, and fluid dynamics. Pump-probe imaging and a streak camera are the most widely used techniques, but they are either demanding the repetitions of the same scene or sacrificing the number of imaging dimensions. Many interesting single-shot ultrafast imaging techniques have been developed in recent years for recording non-repetitive dynamic scenes. Nevertheless, there are still weaknesses in the number of frames, the number of image pixels, or spatial/temporal resolution. Here, we present a single-shot ultrafast microscopy that can capture more than a dozen frames at a time with the frame rate of 5 THz. We combine a spatial light modulator and a custom-made echelon for efficiently generating a large number of reference pulses with designed time delays and propagation angles. The single-shot recording of the interference image between these reference pulses with a sample pulse allows us to retrieve the stroboscopic images of the dynamic scene at the timing of the reference pulses. We demonstrated the recording of 14 temporal snapshots at a time, which is the largest to date, with the optimal temporal resolution set by the laser output pulse. Our ultrafast microscopy is highly scalable in the number of frames and temporal resolutions, and this will have profound impacts on uncovering the interesting spatio-temporal dynamics yet to be explored.

**Introduction**
The technical advance of generating an ultrashort laser pulse has taken the stroboscopic imaging down to events occurring in the picosecond time scales or shorter. One of the most widely used methods is the pump-probe imaging, where a pump beam pulse stimulates a sample and an accompanying probe pulse reads the changes induced by the pump pulse. The pump-probe techniques have been used to visualize the explosion dynamics of water droplet[1], the biological spectroscopy of melanin[2] and laser beam drilling in PMMA[3]. Since the pump-probe delay should be scanned to map the temporal dynamics, it is necessary to irradiate multiple pump pulses for generating identical stimulations in the sample. Therefore, the conventional pump-probe methods are not fully compatible with non-repeating phenomena such as fluid dynamics[4] and explosion dynamics[5,6]. The demand has grown steadily for recording the entire sequence of dynamics induced by a single pump pulse[7,8].

Spectral encoding of the spatio-temporal information has been one of the earliest developments for a single-shot ultrafast dynamic imaging. For example, a broad spectral bandwidth constituting a single short pulse is dispersed in space, and the spatial information encoded in the spectrum is decoded in the detection process[9]. Similarly, the source spectrum is dispersed in time, and the temporal information encoded in the spectrum is decoded by the wavelength-dependent spatial mapping[10,11]. The benefit of these spectral encoding methods is the clear separation of spatial or temporal information in the detector sensors by their wavelengths, which has resulted in optimal measurement sensitivity. On the other hand, the temporal resolution is lower than that set by the original pulse width, since only a partial bandwidth is assigned to each spatial or temporal segment. Another noteworthy approach is to introduce compressive sensing to the conventional high-speed camera for overcoming its limited field of view.

Specifically, the use of a streak camera could image 1D dynamic scenes because the other dimension in the 2D camera sensor is devoted to recording the temporally sheared information. In the compressive sensing approaches, a 2D dynamic scene is spatiotemporally mixed by means of random binary encoding, temporal shearing, and so on in such a way that the 2D dynamic scene is compressed into a single 2D snapshot. The original 2D dynamic scene was retrieved by solving the inverse problem[12,13]. A similar approach has been taken by using spectral mixing instead of temporal shearing by the streak camera[14]. Because no special preparation of illuminations, such as a train of pulses, is necessary, these methods are well suited for imaging with incoherent light as in conventional photography. On the other hand, only a sparse and relatively simple object can be imaged because the compressive recording makes the problem intrinsically underdetermined.

Ultrafast imaging techniques that maintain the laser pulse's full temporal resolution and yet are applicable to arbitrary complex scenes have been proposed on the basis of multiplexing temporal images in the spatial frequency domain. The laser output pulse is divided into a train of reference pulses by using beam splitters or similar optics, and the propagation angles of individual pulses are made different from one another. Then, the stroboscopic recording of a dynamic scene is achieved by the interference of these reference pulses with the sample pulse[15-17]. Since the carrier spatial frequency varies in the interference between each reference pulse and the sample pulse, time-dependent images could be obtained by the spatial Fourier transform of the recorded 2D interference image. A similar approach uses a train of sample pulses with different orientations of stripe patterns in their intensity as structured illuminations to the scene of interest[18]. The temporal resolution of these approaches is the same as that set by the original laser output pulse since the full spectral bandwidth is used in each pulse. The full camera pixels can be used for the view field as the temporal information is multiplexed in the spatial frequency domain. However, these previous studies required a complex optical system to create a train of pulses with different propagation angles and delays. Therefore, only three or four pulses have been used so far, meaning that the number of temporal snapshots is limited to just a few. For pump-probe imaging, the two sets of pulse trains need to be prepared, one for probing the sample and the other for the reference waves, which makes the system even more complicated.

Here, we developed an ultrafast imaging microscopy based on Time and Spatial-Frequency Multiplexing (TSFM) for taking many temporal snapshots of complex scenes and yet achieve the optimal temporal resolution set by the light source. Our TSFM imaging system took the stroboscopic images by the interference of a sample wave with a train of 14 reference pulses that have various propagation angles and time delays. We constructed a unique optical layout composed of a spatial light modulator (SLM) and a custom-made echelon window. A 2D diffraction grating pattern was written on the SLM to split the laser pulse into multiple pulses with different propagation angles. The echelon window added a designed temporal delay to each pulse. The proposed layout is simple, but effective in generating so many reference pulses that the conventional beamsplitter-based approaches can't keep up with them. Using the developed system, we demonstrated the phase and amplitude imaging of a single 100-fs pulse propagating through a turbid medium at 5-THz frame rates. We also visualized the femtosecond scale dynamics of a laser ablation for the successive irradiation of laser pulses. In addition, we realized a single-shot pump-probe shadowgraph imaging of the laser-induced plasma string in air. To make the measurement system simple and scalable in the number of snapshots, a temporally stretched probe beam was prepared to cover the entire dynamics, instead of using a train of pulses whose temporal delays are matched to those of the reference pulses. Our system features the largest number of temporal snapshots demonstrated in the single-shot pump-probe experiments developed so far.

## Results
### Principle of TSFM imaging
To capture the complex-field map of a dynamic scene, $E_S(\boldsymbol{r}, t)$, occurring at the object plane $\boldsymbol{r} = (x, y)$ by the TSFM method, we use specially prepared reference wave composed of $n$ number of multiple pulses with different delays $\tau_j$ and propagation directions $\boldsymbol{k}_R^j$, which is denoted by

$$E_R(\mathbf{r},t) = \sum_{j=1}^{n} a(t-\tau_j)e^{-i\mathbf{k}_R^j \cdot \mathbf{r}}. \tag{1}$$

Here $a(t-\tau_j)$ is the temporal envelop of each reference pulse whose width is set by the laser output pulse width in vacuum. Therefore, the total electric field arriving at the camera located at $z = 0$ is given by

$$E(\mathbf{r},t) = E_S(\mathbf{r},t) + \sum_{j=1}^{n} a(t-\tau_j)e^{-i\mathbf{k}_R^j \cdot \mathbf{r}}. \tag{1}$$

For simplicity, we assumed unity magnification from the object plane to the camera plane. The camera records the intensity of the electric field during the exposure time, which is much longer than the pulse width of the individual waves. Therefore, the signal recorded at the camera can be written as

$$I(\mathbf{r}) = \int_0^\infty \left| E_S(\mathbf{r},t) + \sum_{j=1}^{n} a(t-\tau_j)e^{-i\mathbf{k}_R^j \cdot \mathbf{r}} \right|^2 dt. \tag{2}$$

Then the interference term is given by

$$I_{AC}(\mathbf{r}) = \sum_{j=1}^{n} \int_0^\infty E_S(\mathbf{r},t) a(t-\tau_j) e^{-i\mathbf{k}_R^j \cdot \mathbf{r}} dt + c.c. \tag{3}$$

The electric field of the dynamic scene $E_S(\mathbf{r},t)$ is temporally integrated for the duration of the reference field, i.e. $\tau_j - \Delta t \leq t \leq \tau_j + \Delta t$, for each reference pulse, which is denoted as $\hat{E}_S(\mathbf{r},\tau_j)$. The spatial-frequency spectrum of the interference term, $\tilde{I}_{AC}(\mathbf{k})$, obtained by the 2D Fourier transform of $I_{AC}(\mathbf{r})$ is given as

$$\tilde{I}_{AC}(\mathbf{k}) = \sum_{j=1}^{n} \tilde{E}_S\left(\mathbf{k} - \mathbf{k}_R^j, \tau_j\right) + c.c. \tag{4}$$

Here $\tilde{E}_S$ is the spatial Fourier transform of $\hat{E}_S(\mathbf{r},\tau_j)$. By setting the magnitude of $\mathbf{k}_R^j$ larger than the spatial-frequency bandwidth of $\hat{E}_S(\mathbf{r},t)$ in the detector plane, the spectrum at each time of $\tau_j$ is separated in the frequency domain. After selecting each spectrum $\tilde{E}_S\left(\mathbf{k} - \mathbf{k}_R^j, \tau_j\right)$ associated with the $j^{th}$ reference wave and taking its Hilbert transform, $\hat{E}_S(\mathbf{r}, t = \tau_j)$ is obtained.

**TSFM imaging system**

The experimental setup for TSFM imaging system for imaging the dynamic evolution of a sample pulse gated by the multiple reference pulses is shown in Fig. 1a. A regenerative amplified Ti:Sapphire laser system (Coherent Libra-He, center wavelength: 800 nm, repetition rate: 10 kHz, and pulse width: 91 fs at full width at half maximum (FWHM)) was used as a light source. Only a single pulse was generated at a time using the gated programmable trigger. The output pulse was split by a beam splitter (BS1) and polarizing beam splitter (PBS1) into sample (or pump), reference, and probe beams. The sample or probe beam collected via the objective lens (OL, Olympus RMS 4X, 0.1 NA) was relayed to a camera (sCMOS camera, PCO. Edge 4.2) after recombined with the reference beam by a beam splitter (BS2) to form a Mach-Zehnder interferometry. The magnification from the sample plane to the camera was 13.9, and the field-of-view (FOV) was $515 \times 515\ \mu m^2$ with the diffraction-limited spatial resolution of 4 μm. Scanning mirrors (SM1 and 2) were installed in the sample and probe arms to adjust the overall path-length differences among sample (or pump), probe, and reference beam paths.

To prepare multiple reference pulses with different delays $\tau_j$ and propagation directions $\mathbf{k}_j^R$, we used an SLM (Fourth dimension display, QXGA-R9) and a custom-made glass echelon set in the path of the reference beam. The reference pulse was diffracted into different directions, $\mathbf{k}_R^{(m,n)}$ by a 2D diffraction grating displayed on the SLM. The grating pattern was comprised of a 2D square array of 7-μm size squares with a pitch of 49 μm. Here, $m$ and $n$ indicate the diffraction order along the vertical (Y) and horizontal (X) directions, respectively. The diffracted reference pulse was then focused by a lens (L1) in such a way that the distance between adjacent diffraction spots is 4.0 mm at the focal plane of the lens. The 2D echelon made of a stack of coverglasses (Fig. 1b) was placed on the focal plane of L1, and each diffracted pulse was introduced to each part of the echelon with different stacking number of coverglasses. In this way, the time delay between the neighboring pulses was set to about 200 fs, which is given by the thickness and refractive index of the coverglass. Typically, we chose 14 diffracted pulses with diffraction orders $m = 1, 2$ and $-3 \leq n \leq 3$ in the experiments. Figure 1b indicates the way diffraction order $(m, n)$ was assigned to each segment in the echelon. For simplicity, we assigned the index $j$ to each $(m, n)$ in such a way that the number of stacked coverglasses is increased with the increase of $j$. Therefore, the temporal delay of $j^{\text{th}}$ pulse $\tau_j$ is increased with the increase of $j$. The pulse that travels through one layer of coverglass corresponds to $j=1$, and its flight time was set as a reference point, i.e. $\tau_{j=1} = 0$. Another lens (L2, f=200 mm) delivered these reference pulses to the camera to produce off-axis interference with the sample or probe pulse. In order to calibrate the temporal spacing between the reference pulses, we recorded the total intensity of interference signals between reference pulses and a sample pulse as a function of time delay between the reference and sample arms by scanning the scanning mirror (SM1). Figure 1c shows 14 reference pulses whose relative time delays were measured to be $(\tau_1, \cdots, \tau_{14}) = $ (0, 214, 441, 641, 842, 1015, 1215, 1436, 1623, 1810, 2011, 2197, 2356, and 2555) fs. The intensities of the reference pulses were different depending on the diffraction orders, which was accounted for in Fig. 1c and in the image processing step. The width of each pulse was measured to be about 150 fs, slightly broader than the original laser pulse width due to the residual dispersions in the optical beam paths.

We performed two different types of experiments. The first type of experiment was imaging the propagation of the sample pulse through a medium, which includes three experiments conducted in the following subsections: single-shot light bullet imaging, light propagation through a dynamic scattering medium, and ultrafast videography of glass ablation. In these experiments, a sample pulse was sent across a medium through the sample arm, and a TSFM image of scattered waves from the medium was recorded by the camera via the side-view objective lens and relay lenses. The second type of experiment was the single-shot pump-probe shadowgraph imaging. As an exemplary application, time-resolved transmission shadowgraphs of a laser-induced plasma in air were recorded. In this experiment, a strong pump pulse was tightly focused in air in order to generate a plasma string. A probe pulse sent from the probe arm was transmitted through the plasma string and delivered to the camera via the objective lens. The TSFM shadowgraph images of the plasma string were then recorded.

**Single-shot light bullet imaging**
We demonstrated a single-shot imaging of the trajectory of the light bullet propagating through a weakly scattering medium. A single sample pulse was focused by a spherical convex lens (L3, f = 200 mm) to the water bath, where milk was added at the volume ratio of milk: water = 0.1:1. The sample waves scattered by the particles of fat and protein in milk were captured by the objective lens, as shown in Fig. 2a. The single-shot raw interference image taken during the camera's exposure time of several hundred microseconds is shown in Fig. 2b. Since the light bullet reaches the end of the field of view in just 2 ps, much shorter than the camera exposure time, the raw image showed the trace of the beam path all the way to the bottom of the view field. The inset shows the magnified image indicated by a white dashed box, where a 2D stripe pattern was made visible by the interference between the sample and multiple reference waves. Figure 2c shows the Fourier transform image of Fig. 2b, where many circular spectra were visible because of the multiple reference pulses. Most of the spectra located near the center were attributed to the intensity spectra of sample and reference pulses by themselves, and the interference among reference pulses was also visible. The spectra we are interested in are the interference between the sample beam and reference pulses, which are indicated as red dashed circles along with their associated flight times $\tau_j$. We cropped the spectrum corresponding to each flight time, moved the center of the spectrum to the origin, and took its inverse Fourier transform to obtain the stroboscopic sample image at the designated flight time. Figure 2d shows a set of images obtained for each $\tau_j$ indicated above each sub-image. We could clearly obtain the trajectory of the light bullet, which was propagating from top to bottom. The average spatial intervals of pulses between the snapshots was 45 μm, which corresponds to the travel time of light of about 200 fs in water. This agrees well with the temporal pulse separation measurements in Fig. 1c. The light bullet size measured by the full-width at half maximum of the snapshots in Fig. 2d along the propagation direction about 45 μm, which is equivalent to 200 fs in time. This was broader than the original pulse width due to the pulse dispersion within the sample and optical beam paths.

**Light propagation through a dynamic scattering medium**
We took a single-shot ultrafast image for a short-pulse line beam propagating through a turbid medium. To make the line beam instead of a focused beam, the spherical convex lens (L3) was replaced with a cylindrical lens (f = 50 mm). We prepared the turbid medium by mixing water and milk with the ratio of 1:1. Experiments were performed by the same echelon used in Fig. 2. Figure 3a shows a single-shot raw interference image, which was the accumulation of all the light trajectories during the camera exposure. We applied Hilbert transform algorithm to this raw image and obtained 14 time-lapse images (Fig. 3d). The planar pulse front was attenuated in intensity and broadened in width while propagating through a strongly turbid medium. This can be seen clearly in Fig. 3b, which shows the intensity profile along the vertical line in the various snapshots in Fig. 3d. To the best of our knowledge, this is the largest number of temporal snapshots obtained by the spatial-frequency multiplexing approach. This is also the first real-time visualization of a single pulse propagating through a turbid medium, demonstrating that our method is applicable to spatially complex scenes.

Since the turbid medium changes over time, the way a pulse propagates through the medium varies with time. The turbid medium can be considered to be virtually static during the recording of the series of images shown in Fig. 3d because the time for a single pulse to travel through the view field was only 2 ps, much shorter than the dynamics of milk. On the other hand, when the single-shot ultrafast imaging is taken at times longer than the dynamics of milk, microscopic details of speckle fields will be different. To verify this, we sent pulses from a laser at a repetition of 60 Hz and recorded the single-shot ultrafast images at the same rate as the pulse repetition. We thus obtained the series of images shown in Fig. 3d at the repetition rate of 60 Hz. Figure 3e shows magnified images of the area indicated by the square box in Fig. 3d at the flight time of 842 fs from individual single-shot ultrafast recordings. Therefore, the time difference between the successive images in Fig. 3e is 1/60 second. The speckle granules changed with time, and their wavefronts measured by the phase maps (not shown) varied as well. The normalized field correlation of the images in Fig. 3e with respect to the first image is shown as blue dots in Fig. 3c. The correlation value dropped sharply with the decay constant of 33 ms.

As a control experiment, we recorded the same set of single-shot ultrafast images for a static scattering medium made of PDMS (polydimethylsiloxane) mixed with ZnO nanoparticles (see Materials and methods for details). Figure 3f shows the similar images as in Fig. 3e, but for this static scattering medium. As expected, the images looked almost the same and their field correlations stayed at almost unity (red dots in Fig. 3c).

**Ultrafast videography of glass ablation**
Laser ablation is a representative non-repetitive process. For the proper understanding of the ablation dynamics, it is necessary to record the temporal evolution of a single pulse in the femtosecond timescale. However, the irradiation of a single pulse may not fully ablate the medium, and multiple pulses may need to be irradiated to the sample to complete the process. In order to record the multiple consecutive ablation processes induced by individual pulses in femtosecond timescale over a time span of ~2 ps, strong pump pulses were sent through the sample arm at a rate of 60 Hz and were focused onto the soda-lime slide glass by the spherical convex lens (L3, f = 50 mm). The energy of each pump pulse was about 200 μJ. The first column in Figs. 4a-d show the raw interferograms taken by the camera at 60 Hz. Only four representative images are shown here, and the full movie is available at the supplementary movie. We could notice that the pulses were initially confined at the surface, but the successive irradiation of laser pulses gradually ablated the glass and propagated deeper into the substrate. In this process, the portion of the glass hit by the laser pulses was gradually denatured and acted as an effective waveguide for the successively incoming laser pulses. After the irradiation of 800 pulses, the glass substrate was drilled down to 350 μm.

For each raw image shown in the first column of Figs. 4a-d, we could expand it by the Hilbert transform to reveal temporal snapshots at 200 fs intervals as shown in the images from the second to sixth columns. These femtosecond-scale snapshots visualized the way each pulse was propagating down the glass substrate and inducing the laser ablation. This confirms that our system can investigate ultrafast dynamic scenes over long period of time.

**Single-shot pump-probe shadowgraph imaging**
So far, we have demonstrated the femtosecond-scale imaging of sample pulses themselves as they propagated its way through the sample. Here, we realized a single-shot multi-frame imaging in the pump-probe shadowgraph geometry, where the pump-induced transient transmission changes the propagation of the probe pulses. In the present implementation, we utilized a single time-stretched probe pulse, instead of using a train of multiple probe pulses widely used in the previous studies[15,16]. This makes the experimental setup simple and easily scalable in the number of snapshots. We observed the initial process of generating a plasma string by the irradiation of a strong focused pump pulse to the air molecules. Figure 5a shows the schematic geometry of the pump-probe shadowgraph imaging experiments. A single pump pulse was focused by an objective lens (L3, Olympus RMS 20X, 0.4 NA) to air. The energy of the pump pulse was 300 μJ. After passing twice through a 5-cm thick SF11 glass to stretch its temporal width to ~ 2.3 ps, the probe pulse was transmitted through the plasma string. The temporally stretched probe pulse, imprinted with femtosecond-scale dynamic scenes of the plasma string over a time span of its temporal width, interfered with the train of TSFM reference pulses at the camera plane. The single-shot raw interference image taken by the camera is shown in Fig. 5b. Because the probe pulse was stretched in time, this image contains the accumulated dynamics of the plasma string for ~ 2.3 ps. Temporal snapshots were recovered at times set by the echelon in the same way as in the above experiments. We could expand it by the Hilbert transform to reveal temporal intensity and phase maps at 200-fs intervals as shown in Figs. 5c and 5d, respectively. It is noteworthy that the dynamics are more clearly visible in the phase maps. The pump pulse has not yet arrived at the first snapshot, and then the intensity of the plasma string was gradually increased until the initial 1000 fs. The self-focusing phenomenon was visible after 1200 fs. The size of plasma string was reduced, and its image contrast was increased in the amplitude maps in Fig. 5c. This experiment showed that our reference pulse train scheme could be used to the single-shot pump-probe experiment with great scalability in the number of frames. This will be useful in observing the ultrafast dynamics for a long duration of times with fine temporal steps.

**Discussion**
We proposed a single-shot ultrafast microscopy that can capture 14 snapshots at 200-femtosecond frame interval without loss of the temporal resolution of the original laser pulse. A spatial light modulator and a custom-made echelon were used to generate a large number of reference pulses having different delays and propagation angles. The interference of the sample wave with these reference pulses led to the spatial multiplexing of the temporal images in a single-shot 2D image, from which the stroboscopic images of a dynamic scene were retrieved. We visualized a single short pulse propagating through a turbid medium as it developed spatially complex amplitudes and phase maps. Also, we demonstrated the observation of the femtosecond-scale laser ablation process for the successive irradiation of ultrashort pulses. Furthermore, we demonstrated a single-shot pump-probe imaging by using a single temporally stretched probe pulse and the multiple reference pulses. Our method is highly scalable in the number of temporal snapshots retrievable by a single-shot camera recording and yet achieves the optimal temporal and spatial resolutions. The number of the temporal snapshots can be increased beyond 14 by stacking more coverglasses in the echelon. And the temporal resolution can be flexibly adjusted by embedding the echelon in the liquid medium having different refractive index from air. Therefore, our proposed method will serve as a versatile tool for investigating various non-repetitive dynamic phenomena requiring both high temporal and spatial resolutions.

**Methods and Materials**

**Preparation of static scattering medium**

To make a scattering sample, we added 12 g of ZnO powder to 100 ml of PDMS solution. For a uniform thickness across the view field, we placed a 1.5-ml mixed solution in a 150-mm Petri dish and rotated it at 500 RPM with a spin coater. This produced a 100-µm-thick scattering layer.

**Acknowledgements**

This research was supported by IBS-R023-D1.

**Conflict of interest**

The authors declare that they have no conflict of interest.

**Contributions**

J.M., S.Y. and W.C. conceived the experiment, and J.M. conducted the measurements. Experimental data were analyzed by J.M., S.Y. and W.C. Y.L assisted in the optical set-up design and data analysis. J.M., S.Y. and W.C prepared the manuscript, and all authors contributed to finalizing the manuscript.

**Figure 1**

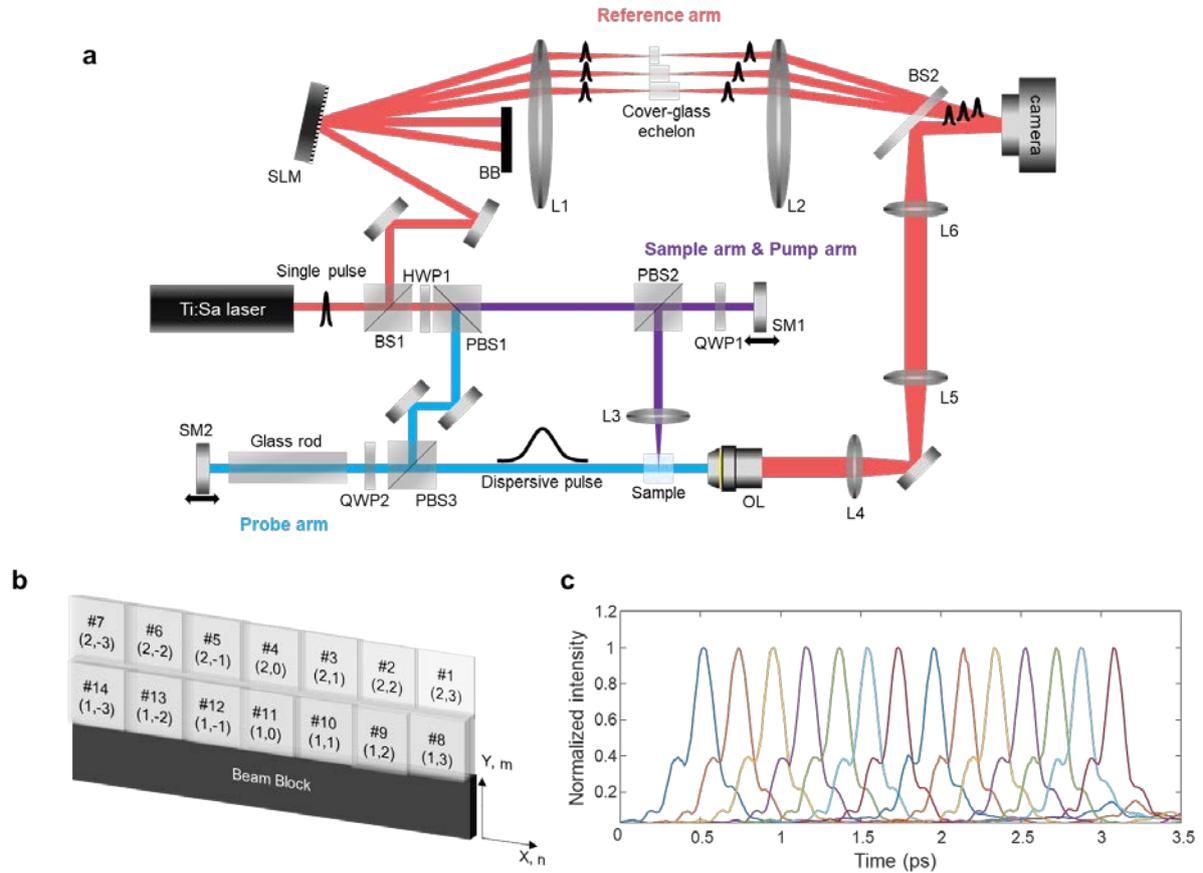

**Figure 1.** Experimental setup for the single-shot ultrafast imaging. (**a**) Schematic diagram of the imaging setup. Reference arm, sample/pump arm, and probe arm are indicated as red, purple, and cyan for visibility although their wavelengths are identical. HWP, half wave plate; QWP1 and 2, quarter-wave plates; PBS1, 2, and 3, polarizing beam splitters; SLM, spatial light modulator; L1-6, lenses; BB, beam block; BS1 and 2, beam splitters; SM1 and 2, scan mirrors; OL, objective lens. (**b**) Echelon diagram. On each segment, the number of coverslips (index $j$ in the main text) and the diffraction order ($m$, $n$) of the diffracted beam by the SLM impinging to the corresponding segment are indicated. (**c**) Interference intensity of multiple reference beams with respect to the single sample pulse taken while the path length of the sample beam was scanned by the SM1. All intensity profiles were normalized by their maximum intensity.

**Figure 2**

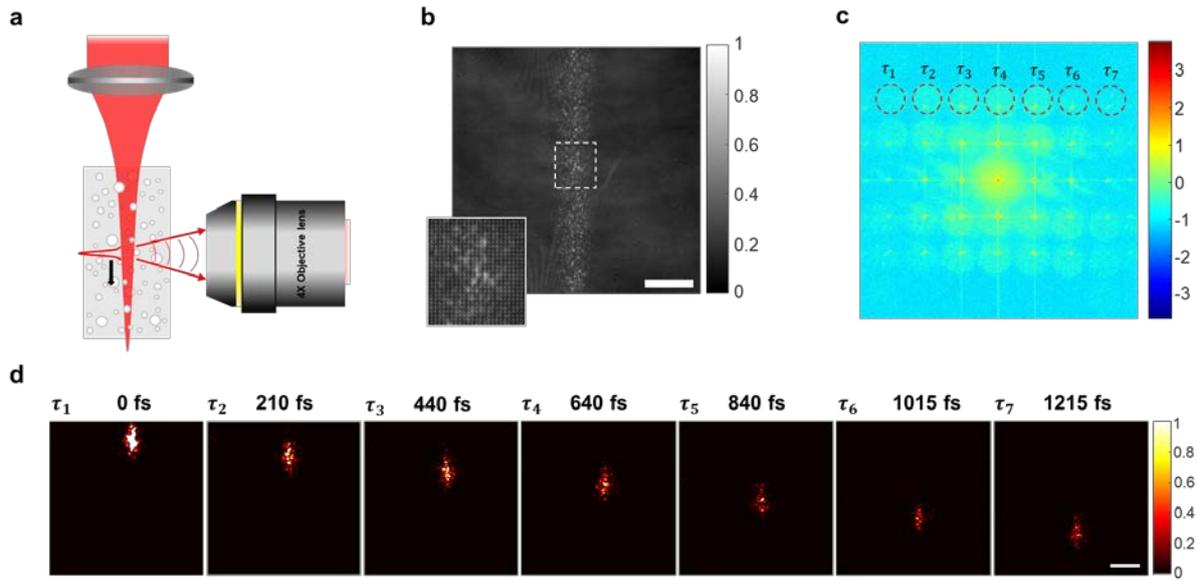

**Figure 2.** Single-shot acquisition of multiple temporal images. (**a**) Schematic geometry of the imaging setup capturing the scattered photons of the laser pulse propagating through a weakly scattering medium made of milk. (**b**) Raw interferogram image. Zoom-in image of the dashed square is shown in the inset, where interference patterns are visible. Scale bar, 100 μm. (**c**) Fourier transform of **b**, where each circle indicated by $\tau_j$ corresponds to the $j^{\text{th}}$ temporal frequency spectrum with the carrier frequency $k_R^j$. (**d**) Temporal images reconstructed by the Hilbert transform of the circular spectra in **c**. We can capture the light bullet at the time interval of 200 fs on average. Scale bar, 100 μm. Color bar, intensity normalized by the maximum intensity in $\tau_1$.

Figure 3

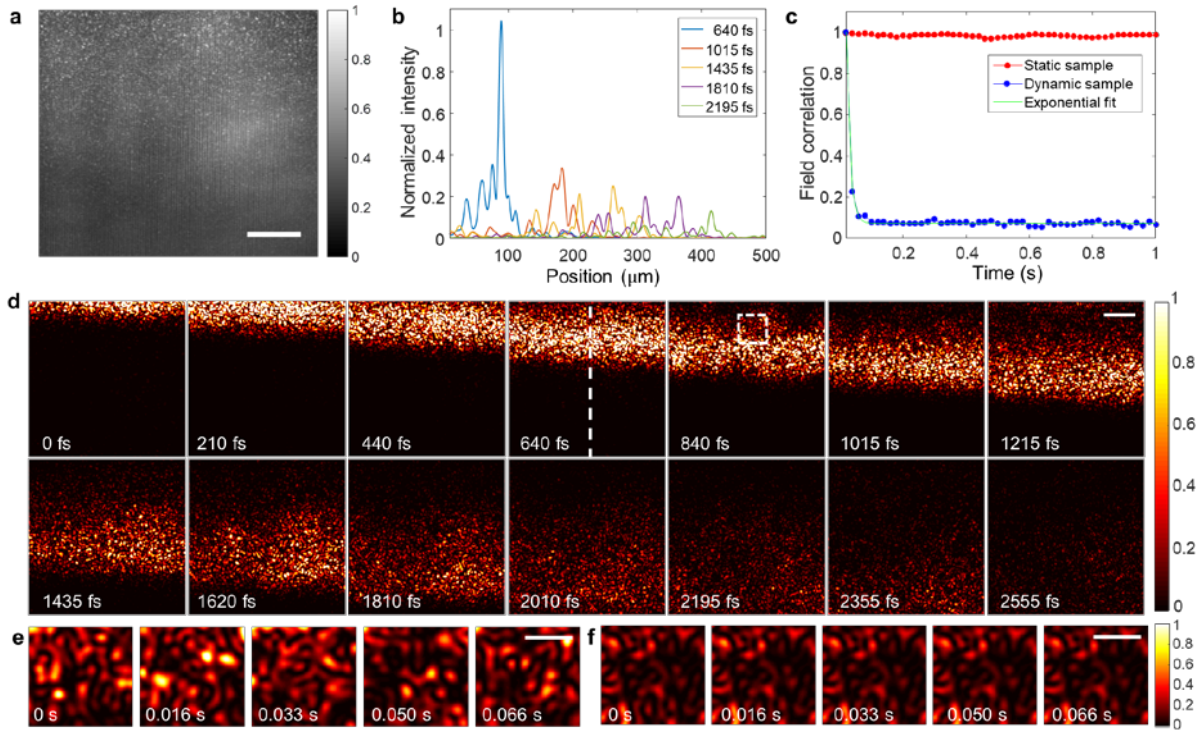

**Figure 3.** Light propagation through static and dynamic scattering media. (**a**) Raw interference image of a line beam propagating through a turbid medium made of milk. Color bar, intensity normalized by the maximum intensity in the image. (**b**) Intensity profiles plotted along the vertical lines shown in **d** for a few representative flight times. Intensity normalized by the maximum intensity in the first profile. (**c**) Normalized field correlation of the images in dynamic sample **e** with respect to the first image (blue dots). Red dots were derived from static sample **f**. Green curve: exponential curve fitting with the temporal decay time of 33 ms. (**d**) Intensity map of temporal images acquired from **a**. Recording time is indicated in each sub-image. Color bar, intensity normalized by the maximum intensity in the first image. (**e**) Zoom-in image of the white dashed box in d at $\tau_{j=5} = 840\ fs$ when single-shot images in **a** were recorded at 60 Hz. (**f**) Same as **e**, but for a static scattering medium made of scattering particles in PDMS. All the scale bars: 100 μm.

**Figure 4**

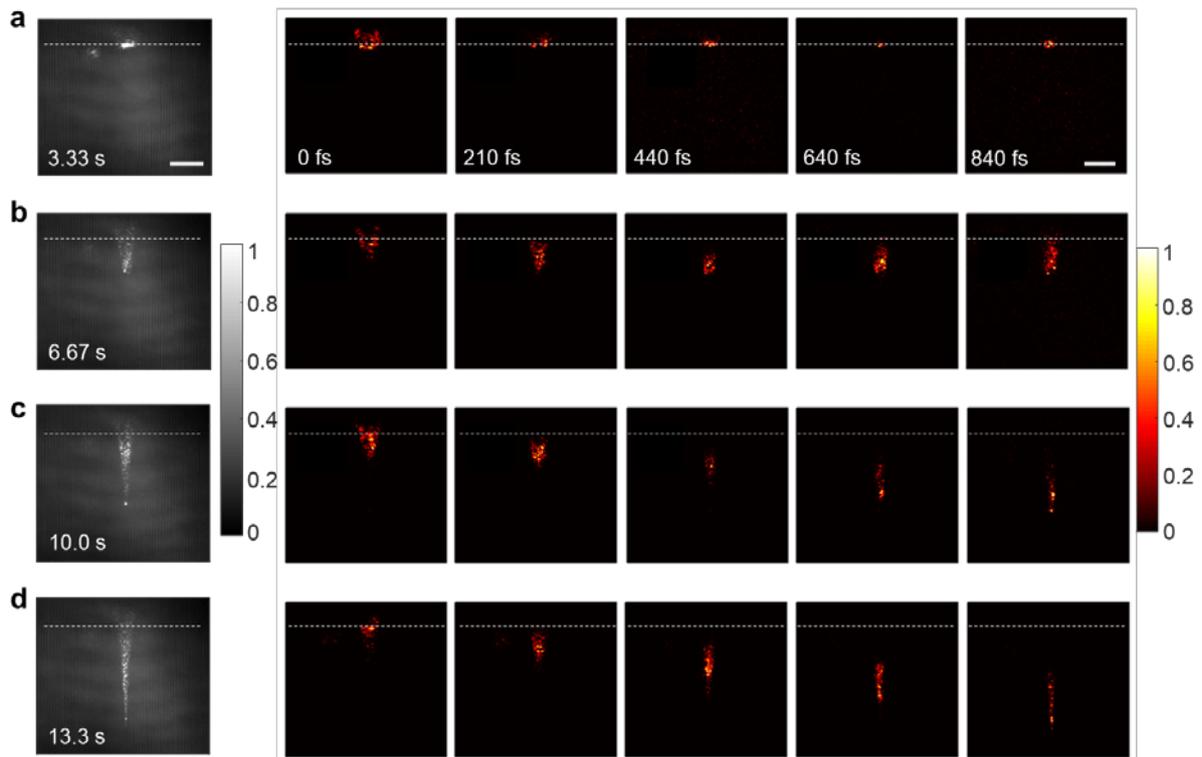

**Figure 4.** Ultrafast videography of the glass ablation by the successive irradiation of ultrashort pulses. (**a-d**) Laser ablation of a glass substrate by the irradiation of four representative laser pulses, i.e. 200th, 400th, 600th and 800th sample beam pulses, among 800 pulses successively illuminated at the rate of 60 Hz. The first column shows the raw interference image, and the images from the second to sixth columns were the snapshots derived from the raw image. A dotted line indicates the surface of the glass. All scale bars, 100 μm.

**Figure 5**

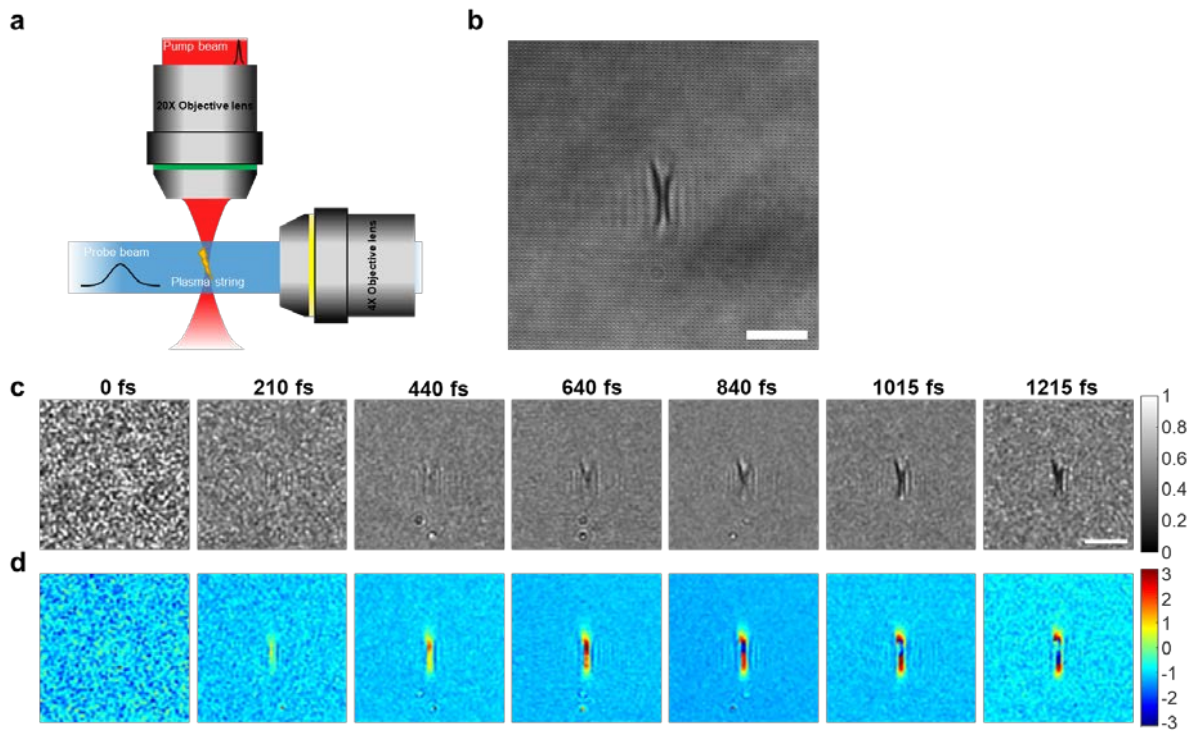

**Figure 5.** Single-shot pump-probe ultrafast imaging of a plasma string in the air. (**a**) Schematic geometry for a pump-probe shadowgraph. (**b**) Raw interferogram image of the plasma string. Scale bar, 100 μm. (**c**) Intensity map of temporal images acquired from **b**. Color bar, intensity normalized by background image taken without the pump beam. Scale bars, 50 μm. (**d**) Phase map of the temporal image acquired from **b**, Color bar, phase in radians.